\documentclass[aps,twocolumn,showpacs,prb]{revtex4}
\usepackage{epsfig}

\begin{document}
%

\title{Selective Spin-State  Switch and Metal-Insulator  Transition in
\boldmath $\rm GdBaCo_2O_{5.5}$.}%
\author{C.      Frontera$^1$,     J.L.      Garc\'{\i}a-Mu\~{n}oz$^1$,
A. Llobet$^2$,
and M.A.G. Aranda$^3$}
\affiliation{$^1$Institut  de Ci\`encia de  Materials de  Barcelona, CSIC,
Campus de la UAB, E-08193 Bellaterra, Spain.\\
$^2$Laboratoire  de  Magnetisme Louis  N\'{e}el,  CNRS  25 Avenue  des
Martyrs - BP 166, 38042 Grenoble Cedex 9, France.\\
%
%
$^3$Departamento de Qu\'{\i}mica Inorg\'{a}nica, Cristalograf\'{\i}a y
Mineralog\'{\i}a Universidad de M\'{a}laga, 29071 M\'alaga, Spain.}

\date{\today}

\begin{abstract}

Ultra-high   resolution   synchrotron   diffraction  data   for   $\rm
GdBaCo_2O_{5.5}$ throw new light  on the metal-insulator transition of
Co$^{3+}$ Ba-cobaltites.  An  anomalous expansion of CoO$_6$ octahedra
is observed at the phase transition on heating, while CoO$_5$ pyramids
show  the   normal  shrinking  at   the  closing  of  the   gap.   The
insulator-to-metal transition is attributed  to a sudden excitation of
some electrons in  the octahedra ($t_{2g}^6$ state) into  the Co $e_g$
band (final $t_{2g}^4e_g^2$ state).   The $t_{2g}^5e_g^1$ state in the
pyramids does  not change  and the structural  study also rules  out a
$d_{3x^2-r^2}/d_{3y^2-r^2}$ orbital ordering at $T_{MI}$.

\end{abstract}

\pacs{71.30.+h,71.38.+i,75.30.Kz}

\maketitle


Transition metal oxides with perovskite structure have demonstrated to
present  a wide variety  of challenging  phenomena.  Superconductivity
and  colossal magnetoresistance of  Cu and  Mn oxides  are spectacular
phenomena  related with  the  strong spin-charge-lattice  correlations
occurring in  these materials.   Cobaltites are also  very challenging
since,  in addition,  the  spin state  degree  of freedom  of Co  ions
introduces    new    effects    in    these   narrow    band    oxides
\cite{Thornton86,Asai98}.  The competition between crystal field (CF),
on-site  Coulomb correlations and  the intra-atomic  exchange energies
leads to the existence of  three possible spin states of $\rm Co^{3+}$
ions: the low spin  state (LS, $t_{2g}^6e_g^0$), the intermediate spin
state   (IS,  $t_{2g}^5e_g^1$),   and   the  high   spin  state   (HS,
$t_{2g}^4e_g^2$).  HS  (LS) is associated with small  (high) values of
CF  (when compared  with the  intra-atomic exchange  energy).   The IS
state appears when  similar values of these two  energies are combined
with the electron-phonon coupling  and the Jahn-Teller distortion that
lifts  the  degeneracy  of  $e_g$  and  $t_{2g}$  orbitals.   In  many
cobaltites the  energy differences between  spin states are  small and
can be easily  overcome by thermal fluctuations and/or  changed by the
lattice  thermal   evolution  \cite{Asai98}  leading   to  spin  state
transitions.

More  recently, a  great  interest on  $\rm LnBaCo_2O_{5+\delta}$  (Ln
$\equiv$      Rare       earth)      cobaltites      has      appeared
\cite{Zhoug94,Troyanchuk98,Maignan99,Suard00a,Vogt00,Suard00b,Moritomo00,Respaud01}.
This family of compounds presents very interesting phenomena like spin
state    transitions    \cite{Suard00a,Moritomo00,Respaud01},   charge
ordering    \cite{Vogt00,Suard00b},   and    giant   magnetoresistance
\cite{Troyanchuk98,Respaud01}.    The  oxygen  content   $\delta$  ($0
\leq\delta\leq 1$) controls the nominal valence of Co ions that varies
from 3.5+ (50\%  of Co$^{3+}$ and 50\% of Co$^{4+}$)  for $\delta = 1$
to 2.5+  (50\% of Co$^{3+}$  and 50\% of  Co$^{2+}$) for $\delta  = 0$
passing through  100\% of  Co$^{3+}$ for $\delta$  = 0.5.   The oxygen
vacancies  introduced when  $\delta <  1$ are  found to  be  placed in
layers together with  the rare earth ions.  Thus,  these compounds are
formed by the  stacking sequence $\rm [CoO_2][BaO][CoO_2][LnO_\delta]$
along the $c$  direction, and Co presents coexistence  of two types of
coordination environments:  pyramidal CoO$_5$ and  octahedral CoO$_6$.
Moreover,  oxygen  vacancies,   within  the  [LnO$_{\delta}$]  layers,
present a remarkably strong  tendency to order. This tendency prevents
the appearance of disorder in the magnetic superexchange interactions,
which    causes   spin    glass   behavior    in    oxygen   deficient
La$_{1-x}$Sr$_x$CoO$_{3-\epsilon}$ compounds.  For  $\delta = 0.5$ the
oxygen   atoms  and   vacancies  are   located  in   alternating  rows
\cite{Maignan99,Moritomo00}.   A  HS  state  has  been  attributed  to
Co$^{3+}$  in pyramidal  environment (a  rather  unusual coordination)
from neutron data in $\rm HoBaCo_2O_5$ (exclusively pyramidal $P\,mma$
structure)  \cite{Suard00b}.    In  contrast,  for   pseudocubic  $\rm
LaCoO_3$, a  large cubic-CF splitting stabilizes,  at low temperature,
the  LS  configuration  \cite{Wu00}  while  at  high  temperature  the
coexistence of  HS and IS  state has been reported  \cite{Asai98}.  LS
has been  proposed for  $\rm LaBaCo_2O_6$ (octahedral  environment) at
low  temperature  \cite{Suard00a}.   Here  we  should  note  that  the
deformation  of the pyramid  in $\delta  =0$ $P\,mma$  cobaltites with
small lanthanides \cite{Suard00b,Vogt00}  is meaningfully different to
that observed  in $\delta =0.5$ cobaltites  and intermediate Ln$^{3+}$
size \cite{Moritomo00},  so that the  differences in the  CF splitting
may  give rise  to different  electronic configurations.   Namely, the
same type  of coordination environment may hold  different spin states
depending  on   the  lattice  deformation   and/or  orbital  occupancy
\cite{Wu00}.

Several   LnBaCo$_{{\rm  2}}$O$_{{\rm   5.5}}$  compounds   present  a
metal-insulator  (MI) transition at  a temperature  ($T_{MI}$) ranging
between 280 and 400  K (depending on Ln).  Susceptibility measurements
reveal that coinciding with this transition there is a large change in
the effective paramagnetic moment  of the samples.  This is understood
as  a sudden  increase,  on heating,  in  the spin  state  of Co  ions
\cite{Maignan99,Moritomo00,Respaud01}.  From  magnetic measurements on
several  samples with  different  lanthanides, all  displaying the  MI
transition,  Maignan  {\em   et  al.}   \cite{Maignan99}  suggested  a
coexistence of IS  (pyramids) and LS (octahedra) for  $T < T_{MI}$ and
HS  Co$^{3+}$ at  high  enough temperatures.   Nevertheless, the  spin
state at both sides of the MI transition is still unclear.  A study of
the structural  changes was carried  out for $\rm  TbBaCo_2O_{5.5}$ by
Moritomo {\em  et al.}  \cite{Moritomo00}.  Based  on these structural
data,  they proposed  a spin  state transition  from a  full  IS state
scheme to  the HS state (for  Ln=Tb) in both  pyramidal and octahedral
sites   \cite{Moritomo00}.    For   $T    >   T_{MI}$   a   HS   state
\cite{Maignan99,Moritomo00}  and a  coexistence of  HS/IS  states have
been  proposed   \cite{Respaud01}  by  different   groups.   The  main
conclusion  of Moritomo  {\it et  al.}~is that  the orbital  degree of
freedom  of the  IS  state ($t_{2g}^5e_g^1$)  and the  electron-phonon
coupling  results  in  a  Jahn-Teller  cooperative  distortion  and  a
$d_{3x^2-r^2}/d_{3y^2-r^2}$  type orbital  order (OO)  below $T_{MI}$.
As the origin  of the transition they proposed  a sudden distortion of
the  basal planes,  on  cooling, that  accommodate the  $d_{3x^2-r^2}$
(pyramid)   and    $d_{3y^2-r^2}$   (octahedron)   orbital   occupancy
\cite{Moritomo00}.      However,     the     sample     studied     in
Ref.~\onlinecite{Moritomo00} had impurities and the reported errors in
the Co-O distances were anomalously large.

\begin{figure}
\centerline{\epsfig{file=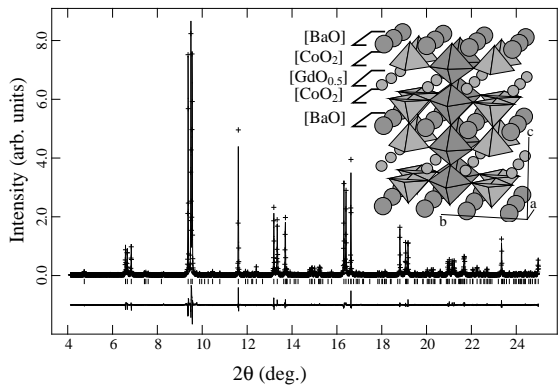,width=0.85\columnwidth}}
\caption{Observed (crosses), calculated  and difference SXRPD patterns
for      $\rm     GdBaCo_3O_{5.5}$      at      $\rm     340\,      K$
($\lambda=0.45029\rm\,\AA$).  The  compound is orthorhombic ($P\,mmm$;
Z=2) with $\rm a_p  \times 2a_p\times 2a_p$ perovskite superstructure.
The  inset shows  a  polyhedral view  of  the orthorhombic  perovskite
structure  of  GdBaCo$_{2}$O$_{5.5}$   (large  circles  are  Ba$^{2+}$
cations and small circles are Gd$^{3+}$ cations).}
\label{fig_dif}
\end{figure}

In  this paper,  we  describe striking  structural  features for  $\rm
GdBaCo_2O_{5.5}$ across $T_{MI}$.   Interatomic Co-O distances exhibit
a  very different evolution  in octahedra  and pyramids.   Analysis of
diffraction   and  magnetic   data   through  the   insulator-to-metal
transition provide evidence for  a low-to-high spin transition only in
octahedra. The $Q_2$-type distortion  of the basal plane of octahedra
and   pyramids  do   not  increase   below  $T_{MI}$   ruling   out  a
$d_{3x^2-r^2}/d_{3y^2-r^2}$ orbital ordering  as the driving force for
the phase  transition. The metallic  phase is caused by  excitation of
electrons, from  Co atoms  placed at the  CoO$_6$ octahedra,  into the
$e_g$ band.

$\rm  GdBaCo_2O_{5+\delta}$  was   prepared  by  standard  solid-state
reaction  in  air  at  $T=1125^{\rm  o}C$  during  $24\rm\,h$.   After
regrinding of the pellet, the compression and annealing processes were
repeated  several times.   As Gd  has a  very high  neutron absorption
coefficient, we have characterized the structural evolution across the
MI transition using  X-ray thermodiffractometry. Ultra-high resolution
synchrotron X-ray  powder diffraction (SXRPD)  patterns were collected
at   BM16  diffractometer   of   ESRF  (Grenoble)   in  the   standard
Debye-Scherrer configuration.   The polycrystalline sample  was loaded
in  a  borosilicate glass  capillary  ($\phi=0.5\rm\,mm$) and  rotated
during    data   collection.     A    short   wavelength,    $\lambda=
0.450294(6)\rm\,\AA$ ($27.54\rm\,keV$), selected with a double-crystal
Si $(1\,1\,1)$ monochromator, and  calibrated with Si NIST ($a=5.43094
\rm\,\AA$), was chosen to  reduce the sample absorption.  Measurements
have been  done at T= 300,  320, 340, 360, 380  and $400\rm\,K$.  Each
SXRPD run took about  $\frac{3}{4}\rm\,h$ to have good statistics over
the  angular  range  $4^{\rm  o}\leq  2\theta \leq  25^{\rm  o}$  with
$0.005^{\rm o}$ step  size.  The powder patterns were  analyzed by the
Rietveld method  using the GSAS suite of  programs \cite{Larson94}. No
impurities  peaks have been  detected and  the diffraction  peaks were
remarkably sharp for a  three metal-containing perovskite.  The oxygen
stoichiometry was determined  to be 5.53(1) from SXRPD  data, an usual
oxygen  content found  for  air-synthesis \cite{Maignan99,Moritomo00}.
Moreover, according  to Ref.~\onlinecite{Maignan99} the  MI transition
in absent in  Gd samples with $\delta \geq  0.6$.  Characterization of
the sample  included magnetic measurements (SQUID)  in the temperature
range $10{\rm\, K}\leq  T \leq 650\rm\,K$.  Magnetotransport, magnetic
and  optical  transmission  measurements  under pulsed  fields  up  to
$35\rm\,T$ have been also carried  out \cite{Respaud01}.

First, we want  to focus on the crystal structure  and the ordering of
the oxygen vacancies.  The structure of $\rm TbBaCo_2O_{5.5}$ reported
in  Ref.~\onlinecite{Moritomo00} was  used as  starting model  for the
Rietveld refinements. We will use the same atomic labelling scheme for
the sake of comparison. In the crystal structure ($P\,mmm$) the simple
perovskite cell is doubled along  the $b$-axis in order to account for
alternating oxygen rich and  oxygen deficient $a-c$ layers. The oxygen
vacancies were  checked in our $\rm GdBaCo_2O_{5.5}$  sample and found
to be located at the lanthanide layers.  O3' at (0,0,$\frac{1}{2}$) is
almost  empty with  a refined  occupation  factor of  0.08(2).  O3  at
(0,$\frac{1}{2}$,$\frac{1}{2}$)  is  fully  occupied  with  a  refined
occupation factor  of 0.99(2).  So,  there are octahedra  chains along
the  $c$-axis alternated  along the  $b$-axis with  the corner-sharing
square  pyramids in both  cobaltites.  Figure~\ref{fig_dif}  shows the
SXRPD Rietveld plot at 340 K as an example. Refined atomic coordinates
and  agreement factors  are given  in Tab.~\ref{tabpos}.   The crystal
structure  of   $\rm  GdBaCo_2O_{5.5}$  is  shown  in   the  inset  of
Fig.~\ref{fig_dif}.   From  this structure  it  is  apparent that  the
metallic  phase  should   be  highly  anisotropic.   Oxygen  vacancies
preclude the existence of highly conducting paths along $c$ within the
$a-c$ layers of pyramids.

\begin{figure}
\centerline{\epsfig{file=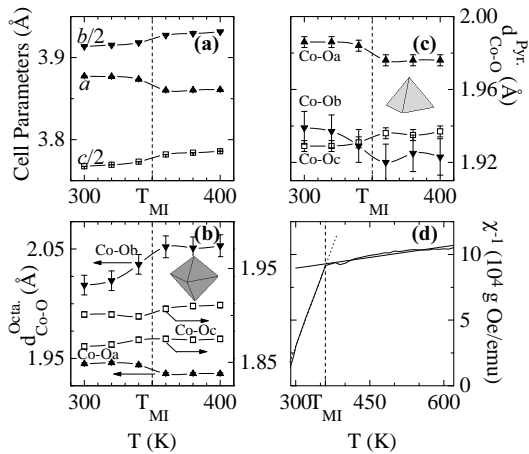,width=0.80\columnwidth}}
\caption{Temperature dependence  of {\bf (a)}  lattice constants; {\bf
(b)} Co-O bond  distances for the CoO$_6$ octahedra  (Co-Oi stands for
the bond lengths along i=$a$, $b$  and $c$ axes); {\bf (c)} Co-Oi bond
distances  for the  CoO$_5$ pyramids;  and  {\bf (d)}  inverse of  the
susceptibility $\chi_{\rm Co}$ (the straight lines show the Curie fits
above and below $T_{MI}$.}
\label{fig_dis}
\end{figure}

On  cooling through  the transition,  $b$ and  $c$  lattice parameters
exhibit a  sudden shrink (0.28\%  and 0.27\%, respectively)  while $a$
lengthens at  $T_{MI}$ (0.35\%), see  Fig.~\ref{fig_dis}(a).   The
values  found, and  their  thermal evolution,  strongly contrast  with
those    reported    in    Ref.~\onlinecite{Moritomo00}    for    $\rm
TbBaCo_2O_{5.5}$, where $a> b/2$, $a$ shrinks (on cooling) at $T_{MI}$
and $b/2$  lengthens.  Figures \ref{fig_dis}(b)  and \ref{fig_dis}(c)
show  the Co-O bond  distances, for  Co1O$_6$ octahedron  and Co2O$_5$
pyramid  respectively, along  the three  crystallographic  axes (Co-Oi
stands for the lengths along i=$a$, $b$ and $c$ axes) derived from the
refined  atomic coordinates  given in  Tab.~\ref{tabpos}.   Again,
there   are   important   differences    with   the   case   of   $\rm
TbBaCo_2O_{5.5}$. We have found that pyramids and octahedra are
deformed in both the insulating  and metallic states. The longest Co-O
distance is Co-Oa for the former (Co2O$_{5}$) and Co-Ob for the latter
(Co1O$_6$).  Long and short bonds alternate along the $b$ axis at both
sides of  the transition.  A  pattern well different to  the evolution
proposed  in Ref.~\onlinecite{Moritomo00}  for  $\rm TbBaCo_2O_{5.5}$,
where  that  alternation was  only  clearly  observed below  $T_{MI}$,
proposing that  the basal plane deformation in  pyramids and octahedra
vanishes      above     $T_{MI}$.      Hence,      conversely     with
Ref.~\onlinecite{Moritomo00}, our data  discard the stabilization of a
$d_{3x^2-r^2}/d_{3y^2-r^2}$-type  orbital ordering  below  $T_{MI}$ as
the physical mechanism for the MI transition in $\rm GdBaCo_2O_{5.5}$.
Furthermore,  we show  in Figs.~\ref{fig_dis}(b)  and \ref{fig_dis}(c)
that the  difference between  long and short  Co-O bonds in  the basal
plane   (the   $Q_2$-type   antiferrodistorsive  distortion)   remains
practically  unchanged in the  pyramid ($0.06\rm\,\AA$)  but increases
strongly  in  the  octahedron   when  heating  above  $T_{MI}$.   This
evolution   clearly  rules  out   the  aforementioned   orbital  order
transition.

Fig.~\ref{fig_dis}(d) shows the  temperature dependence of the inverse
susceptibility (measured  in a field  of $1\rm\,T$ after a  zero field
cooling process)  in the paramagnetic  region up to  $625\rm\,K$.  The
paramagnetic   contribution   from   Gd   atoms   (estimated   between
10-$625\rm\,K$,   $\chi_{\rm  Gd}=   \displaystyle   \frac  {1.92\cdot
10^{-2}}{T+0.4} \frac{emu}{g\,Oe}$) was subtracted in order to extract
the    contribution   coming   only    from   Co    ions   [$\chi_{\rm
Co}=1/2(\chi-\chi_{\rm  Gd})$].   A drastic  change  in the  effective
moment $\mu_{\rm  eff}$ and  the sign of  the Curie  temperature takes
place coinciding  with the  electronic localization. According  to the
Curie-Weiss fitting shown in Fig.~\ref{fig_dis}(d) $\mu_{\rm eff}^{\rm
Co}$ per Co atom changes from $\mu_{\rm eff}^{\rm Co}=1.8(1)\,\mu_{\rm
B}$ ($T<T_{MI}$) to  $4.3(2)\,\mu_{\rm B}$ ($T>T_{MI}$).  The expected
values for Co$^{3+}$ full IS, 1:1  mixture LS/IS, IS/HS or full HS are
respectively 2.83, 2.00, 4.00 and 4.90 $\mu_{\rm B}$.  The possibility
of  a  non-negligible  orbital  contribution  to the  moment  in  this
compound is uncertain \cite{Kwon00}.   However, below $T_{MI}$ (in the
paramagnetic insulating phase) the  effective moment found per Co atom
agrees with  a 1:1 mixture of Co$^{3+}$  in LS and IS,  ruling out the
full Co$^{3+}$ IS  scheme. Since IS Co$^{3+}$ is  most probable in the
pyramids  \cite{Maignan99,Kwon00}, the  results  indicate that,  below
$T_{MI}$,  the octahedron  contains LS  Co$^{3+}$ and  the  pyramid IS
Co$^{3+}$.   Interestingly, the  moment  per Co  atom  deduced in  the
metallic phase below $625\rm\,K$ agrees  with a half of Co$^{3+}$ ions
in IS and the other half in HS states.

\begin{figure}
\centerline{\epsfig{file=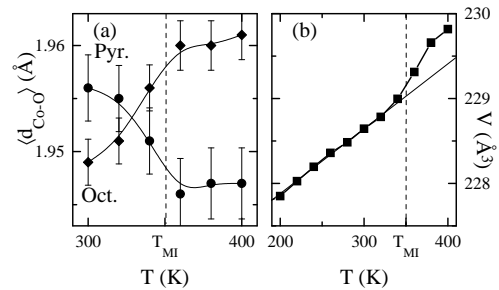,width=0.75\columnwidth}}
\caption{Temperature  dependence of {\bf  (a)} average  Co-O distances
for the  CoO$_{6}$ and  CoO$_{5}$ polyhedra and  {\bf (b)}  unite cell
volume.    The  continuous   straight   line  in   {\bf   (b)}  is   a
guide-to-the-eyes  to  highlight  the  anomalous volume  expansion  at
$T_{MI}$.}
\label{fig_vol}
\end{figure}

Before discussing  the most probable scenario  for the metal-insulator
transition,  let  us analyze  in  more  detail  the Co-O  bond  length
variation.   Does that  structural evolution  corroborate  the picture
deduced  from magnetic  data? The  answer  is positive.   As shown  in
Fig.~\ref{fig_dis}(c), Co-Oa and Co-Ob  basal distances of the pyramid
both lengthen  on cooling, in  a so similar  way that the  basal plane
deformation remains  practically unchanged at both  sides of $T_{MI}$.
Moreover,  the apical  Co-Oc distance  of the  pyramid  changes little
across the  transition.  We want  to emphasize that these  changes are
very different to those observed in the octahedron.  They are shown in
Fig.~\ref{fig_dis}(b) and Fig.~\ref{fig_vol}(a): the Co-Ob distance in
the  octahedra displays a  pronounced shrinking  on cooling,  which is
accompanied by  a moderate increase  of the Co-Oa distance  (again the
Co-Oc apical distances change little).   As a result, in contrast with
the pyramid, the difference between  the two diagonal distances of the
octahedra increases  notably when enters the metallic  phase.  At this
point  it  is  important to  recall  that  a  transition to  a  higher
spin-state  in Co$^{3+}$  (as deduced  from magnetic  data)  implies a
bigger  effective  ionic  radius  which should  lengthen  the  average
$\langle d_{\rm Co-O}\rangle$ bond  length.  Hence, a very significant
finding  is the  different evolution  of the  average  $\langle d_{\rm
Co-O}\rangle$   distance   in   the   octahedron   and   the   pyramid
[Fig.~\ref{fig_vol}(a)].   So, a  central result  is that  the average
$\langle  d_{\rm Co-O}\rangle$  distance of  the  octahedron increases
substantially at  $T_{MI}$ on  heating.  This finding  constitutes the
first experimental result  establishing that the spin-state transition
in $\rm LnBaCo_2O_{5.5}$ occurs solely in the octahedra.  Thus, in the
pyramids  we   observe,  on  heating,   the  metal-oxygen  bond-length
contraction normally found when the gap closes and enters the metallic
phase,  indicating again that  the transition  to a  higher spin-state
detected in the susceptibility takes  place in the octahedron, but not
in the pyramid.  This transition is also responsible for the anomalous
volume  expansion   at  $T_{MI}$  plotted   in  Fig.~\ref{fig_vol}(b).
Normally,  the cell  volume  contracts in  an electron  delocalization
process.  The  volume expansion observed here  is another confirmation
of the transition to a higher spin-state.

\begin{table*}
\caption{Atomic   positions  for  $\rm   GdBaCo_2O_{5.53(1)}$  from
synchrotron powder diffraction Rietveld refinements.  The sites are Ba
2o $(\frac 12,y,0)$;  Gd 2p $(\frac 12,y,\frac 12)$;  Co1 2r $(0,\frac
12,z)$; Co2 2q $(0,0,z)$; O1  1a $(0,0,0)$; O2 1e $(0,\frac 12,0)$; O3
1g  $(0,\frac  12,\frac   12)$;O3'  1c  $(0,0,\frac  12)$  [occupation
factor$=0.08(2)$];  O4 2s  $(\frac  12,0,z)$; O5  2t $(\frac  12,\frac
12,z)$; O6 4u $(0,y,z)$.}
\label{tabpos}

\begin{tabular}{lccccccccc}
$T(K)$ & Ba($y$) & Gd($y$) & Co1($z$) & Co2($z$) & O4($z$) & O5($z$) &
O6($y$) & O6($z$) & $R_F\;(\%)$ \\ \hline
300 &  0.2500(2) &  0.2722(2) & 0.2522(5)  & 0.2561(4) &  0.3132(16) &
0.2737(18) & 0.2450(12) & 0.2929(11) & 3.91 \\
320 &  0.2500(2) &  0.2718(2) & 0.2521(5)  & 0.2559(4) &  0.3134(16) &
0.2745(18) & 0.2449(12) & 0.2925(11) & 3.77 \\
340 &  0.2496(2) &  0.2708(2) & 0.2517(5)  & 0.2558(4) &  0.3131(15) &
0.2744(17) & 0.2434(11) & 0.2940(10) & 3.80 \\
360 &  0.2493(2) &  0.2686(2) & 0.2521(5)  & 0.2560(5) &  0.3119(16) &
0.2723(18) & 0.2418(12) & 0.2936(10) & 4.20 \\
380 &  0.2492(2) &  0.2685(2) & 0.2524(5)  & 0.2557(5) &  0.3115(16) &
0.2720(18) & 0.2421(12) & 0.2939(10) & 4.02 \\
400 &  0.2493(2) &  0.2683(2) & 0.2524(5)  & 0.2558(5) &  0.3116(16) &
0.2723(18) & 0.2419(12) & 0.2934(10) & 4.36
\end{tabular}
\end{table*}

Recapitulating,   the   analysis  of   the   structural  changes   and
susceptibility data for $\rm  GdBaCo_2O_{5.5}$ has permitted to draw a
detailed  picture  of  the  metal-insulator  transition  in  the  $\rm
LnBaCo_2O_{5.5}$ family  of compounds.   Our results reveal  that: (i)
There  is   a  sudden  expansion   of  the  average   $\langle  d_{\rm
Co-O}\rangle^{\rm octa}$ distance  at T$_{{\rm MI}}$, concomitant with
a spin-state transition from LS (insulating) to HS (metallic) state in
the Co$^{3+}$  ions of the octahedra.   (ii) Co atoms  in the pyramids
hold  the  same  spin-state   (Co$^{3+}$  IS)  before  and  after  the
electronic  delocalization.   Hence,  the  pyramid simply  shrinks  as
commonly  observed in  Mott  oxides when  enters  the metallic  phase.
(iii) These findings imply the  existence of spin state ordering below
and above $T_{MI}$.  (iv) The alternation of short and long Co-O bonds
along  the $b$  axis is  present in  the insulating  and  the metallic
phases   (and    not   only   in   the   former,    as   reported   in
Ref.~\onlinecite{Moritomo00}   for    $\rm   TbBaCo_2O_{5.5}$).    The
$Q_2$-type distortion  (antiferrodistorsive) of the  basal planes does
not    increase   in    the   insulating    phase,   ruling    out   a
$d_{3x^2-r^2}/d_{3y^2-r2}$ type orbital ordering  as the origin of the
transition.  To conclude, the driving force for the MI transition is a
spin-state  switch in  the Co$^{3+}$  ions located  at  the octahedra.
They suddenly switch from LS ($t_{2g}^6e_g^0$) to HS ($t_{2g}^4e_g^2$)
state at $T_{MI}$.  Thereby,  the metallic conductivity in this family
of  oxides  (full  Co$^{3+}$)  seems  related with  the  injection  of
electrons in the conduction band that accompanies the stabilization of
a HS ($t_{2g}^4e_g^2$) state in the CoO$_6$ octahedra.

Financial   support  by  the   MEC  (PB97-1175),   CICyT  (MAT97-0699,
MAT97-326-C4-4,  and MAT99-0984-C03-01)  and Generalitat  de Catalunya
(GRQ95-8029)  projects  is  thanked.   ESRF is  acknowledged  for  the
provision of X-ray synchrotron facilities  and Dr. E.  Dooryee for his
assistance during data collection.



\end{document}